\documentclass[%
unsortedaddress,
preprint,
 amsmath,amssymb,
 aps,
prb,
]{revtex4-1}

\usepackage{graphicx}% Include figure files
\usepackage{dcolumn}% Align table columns on decimal point
\usepackage{bm}% bold math
\usepackage{amsmath}
\usepackage{multirow}% Added by H. Shin to plot multi-row table
\usepackage{xcolor}
\begin{document}

\title{DFT+U and Quantum Monte Carlo study of electronic and optical properties of AgNiO$_2$ and AgNi$_{1-x}$Co$_{x}$O$_2$ delafossite}
\author{Hyeondeok Shin}%
\email{hshin@anl.gov}
\affiliation{Computational Science Division, Argonne National Laboratory, Argonne, Illinois 60439, USA}
\author{Panchapakesan Ganesh}
\affiliation{Center for Nanophase Materials Sciences, Oak Ridge National Laboratory, Oak Ridge, Tennessee 37831, USA}
%Authors' institution and/or address\\This line break forced with \textbackslash\textbackslash}%
\author{Paul R. C. Kent}
\affiliation{Computational Sciences and Engineering Division, Oak Ridge National Laboratory, Oak Ridge, TN 37831-6494}
\author{Anouar Benali}
\affiliation{Computational Science Division, Argonne National Laboratory, Argonne, Illinois 60439, USA}
\author{Anand Bhattacharya}
\affiliation{Materials Science Division, Argonne National Laboratory, Lemont, Illinois 60439, USA}
\author{Ho Nyung Lee}
\affiliation{Materials Science and Technology Division, Oak Ridge National Laboratory, Oak Ridge, TN 37831, USA}
\author{Olle Heinonen}\altaddress{Present and permanent address: Seagate Technology, Computer Ave. 7801, Bloomington, MN 55435}
\affiliation{Material Science Division, Argonne National Laboratory, Argonne, Illinois 60439, USA}
\author{Jaron T. Krogel}
\email{krogeljt@ornl.gov}
\affiliation{Materials Science and Technology Division, Oak Ridge National Laboratory, Oak Ridge, TN 37831, United States}


\date{\today}% It is always \today, today,

\begin{abstract}
As the only semimetallic $d^{10}$-based delafossite, AgNiO$_2$ has received a
great deal of attention due to both its unique semimetallicity and its
antiferromagnetism in the NiO$_2$ layer that is coupled with a lattice distortion. In
contrast, other delafossites such as AgCoO$_2$ are insulating. Here 
we study how the
electronic structure of AgNi$_{1-x}$Co$_{x}$O$_2$ alloys vary with Ni/Co
concentration, in order to investigate the electronic properties and phase
stability of the intermetallics.
While the electronic and magnetic structure of delafossites have been studied using Density Functional Theory (DFT), earlier studies have not
included corrections for strong on-site Coulomb interactions. In order to treat
these interactions accurately, in this study we use Quantum Monte Carlo (QMC) simulations to
obtain accurate estimates for the electronic and magnetic properties of
AgNiO$_2$. By comparison to DFT results we show that these electron correlations
are critical to account for. 
We show that Co doping on the magnetic Ni sites
results in a metal-insulator transition near $x\sim 0.33$, and reentrant behavior near $x\sim 0.66$

\end{abstract}

\pacs{Valid PACS appear here}% PACS, the Physics and Astronomy

\maketitle

%\tableofcontents

\section{\label{sec:level1}Introduction}
\begin{figure}[t]
 \includegraphics[width=6 in]{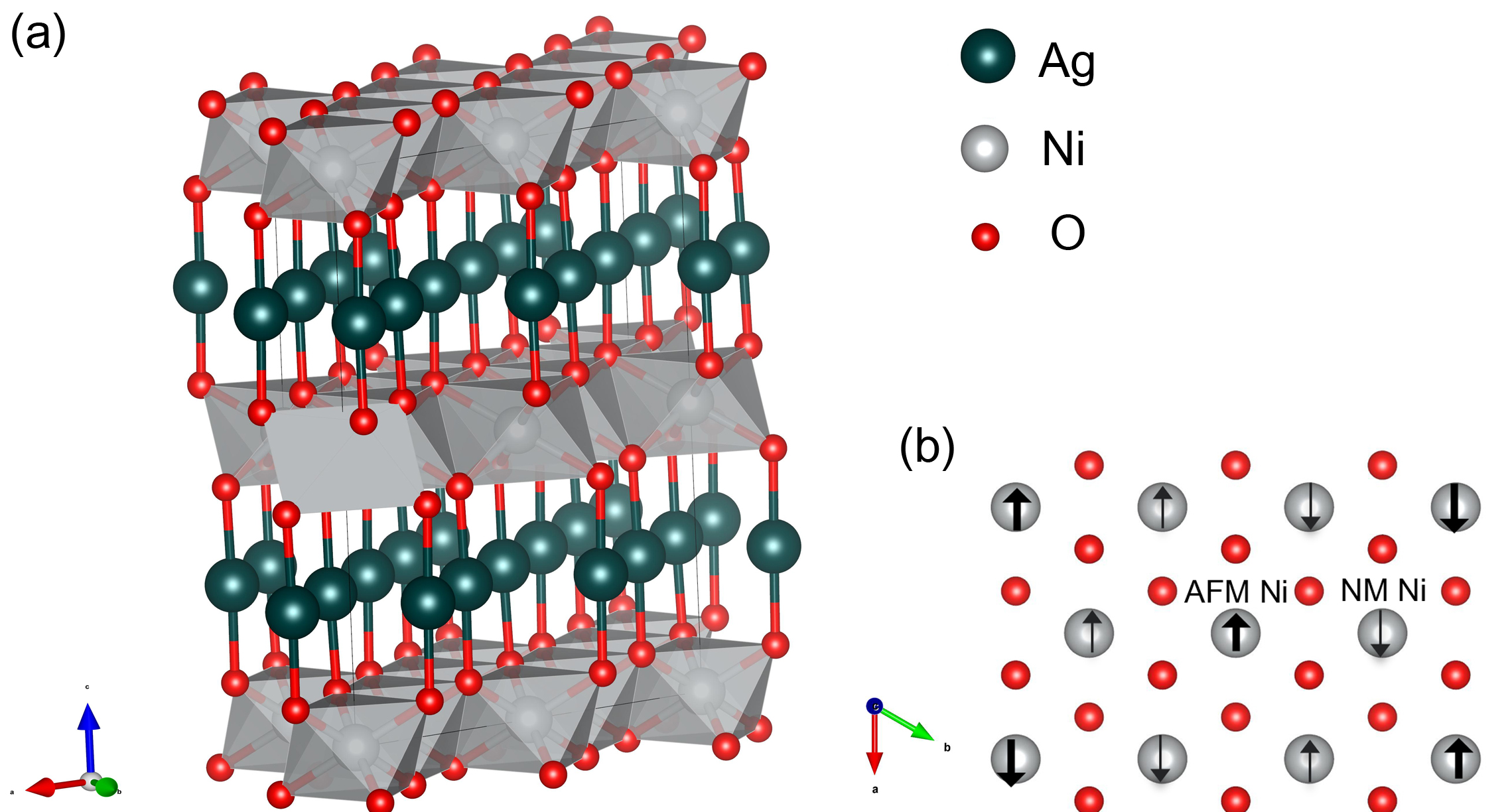}
 \caption{(a) side and (b) top view of 2H-AgNiO$_2$ structure. There are two different Ni sites, Ni1 with very small magnetic moments as Ni$^{+3.5}$/$d^{6.5}$ state, and Ni2 with large magnetic moments as Ni$^{+2}$/$d^{8}$ state; the Ni2 sites form a triangular planar antiferromagnet.}
 \label{fig:AgNiO2}
\end{figure}

Delafossites are minerals with the generic formula ABO$_2$, where A is a monovalent and B is a trivalent metal, and the structure consists of layers of the metal A cations interspersed between layers of BO$_2$ that are arranged in edge-sharing BO$_6$ octahedra, as seen in Figure~\ref{fig:AgNiO2}. A particular feature of delafossites is that the cation A is bonded vertically (along the $c$ axis) to oxygen atoms in planes above and below. Recently, delafossites have attracted a great deal of attention because of their interesting properties that arise as a consequence of interplay between the  metal A and the BO$_2$ layers . Since the discovery of the naturally occurring delafossite form of CuFeO$_2$,\cite{pabst46} various types of delafossites have been synthesized and studied extensively in order to understand what gives rise to the wide ranges of electronic properties for different combinations of A and B elements.~\cite{shannon71,prewitt71,rogers71,benko84,benko87,kawazoe97,noh09,kushiwaha15,ok20}. The monovalent A site is  usually occupied by $d^9$ or $d^{10}$ noble metals or transition metal atoms. Most of the $d^9$ and $d^{10}$ delafossites, such as PdCoO$_2$ and PtCoO$_2$, exhibit a large electrical conductivity, but a much wider range of electronic properties, including metallic, semiconducting, and insulating, has been observed in $d^{10}$ A-site compounds, and it appears that the B-site component is the dominating factor in the resulting electronic and optical properties.\cite{wawzynska07,kang07,seki08}

Delafossites with $d^{10}$ cations (A = Ag and Cu) have reported to possess wide direct electronic band gaps and $p$-type behavior, which makes them interesting for potential future applications of $p$-type transparent materials.~\cite{kawazoe97,ueda01,yanagi01,snure07,scanlon09,santra13} 
Among $d^{10}$-based delafossites AgBO$_2$ and CuBO$_2$, the Ni B-site compound AgNiO$_2$ is known to possess rather unique electronic and magnetic properties: while most of the $d^{10}$-based delafossites exhibit insulating or semiconducting behavior, only AgNiO$_2$ exhibits metallic features in 2H polytype of hexagonal space group of P6$_3/mmc$.~\cite{wichainchai88,wawzynska07,coldea14}
According to previous studies on 2H-AgNiO$_2$, its ideal P6$_3/mmc$ crystal structure is transformed into the P6$_3$22 structure because of lattice distortions induced by strong antiferromagnetic (AFM) interactions in the NiO$_2$ layers of AgNiO$_2$.~\cite{wawzynska07,wawzynska08}
Interestingly, the lattice distortion in AFM 2H-AgNiO$_2$ is not the well-known Jahn-Teller distortion, but a charge-ordering distortion induced by charge transfer on e$_g$ states on the Ni sites. This leads to two different Ni sites, Ni1 with small magnetic moments (Ni$^{3.5+}$), and Ni2 sites with large magnetic moments (Ni$^{2+}$); the Ni2 sites form a triangular antiferromagnet within the Ni $ab$-plane. 
This leads to charge disproportionation on the Ni sites and AFM 2H-AgNiO$_2$ is consequently interpreted as a strongly charge-ordered system.
Moreover, because of the insulating properties of AgCoO$_2$ while AgNiO$_2$ exhibits a semimetallic phase, the existence of a metal-insulator transition has been predicted on AgNi$_{1-x}$Co$_{x}$O$_2$ structures wherein the NiO$_2$ layers in AgNiO$_2$ are mixed with CoO$_2$ layers of the insulating AgCoO$_2$.~\cite{shin93}

In addition to experimental investigations, there have been a few reports from studies using density functional theory (DFT) to study AFM 2H-AgNiO$_2$, in particular to address the magnetic order that has been observed experimentally.~\cite{wawzynska07,pickett14}. A fundamental question that can be raised in this context is to what extent electronic correlations play a role in the magnetic ordering in delafossites in general, and in AgNiO$_2$ in particular; linked to this is the well-known broader issue of how to accurately account for electronic correlations within DFT.  This is an important question for the delafossites as they contain 3d, 4d, 4f, and 5f metals with highly localized electrons bound to oxygen. It is therefore important to accurately assess the effects of electronic correlations on delafossites, and also to devise computational schemes that allow for including correlations at a known level of accuracy. 
One such scheme is DFT+U, in which a Hubbard U term is added to selected localized orbitals to approximately account for on-site Coulomb correlations.~\cite{anisimov91,dudarev98} While the actual value of U can be used as a fitting variable, there are nowadays methods to self-consistently calculate U, reducing empiricism. 
Nevertheless, an on-site Coulomb interaction Hubbard U has not been considered at all in previous DFT studies of 2H-AgNiO$_2$, mainly because it has been predicted that the effect of U is small in metallic 2H-AgNiO$_2$\cite{wawzynska07}. 
Furthermore, a previous DFT study for 3R-AgNiO$_2$ concluded that projected density of state from local spin density approximation (LSDA) is in better agreement with corresponding experimental partial spectral weight (PSW) distributions than LSDA+U.~\cite{johannes07}    
Therefore, appropriate values of U for 2H-AgNiO$_2$ have not been  studied systematically. Previous studies have utilized the local density approximation (LDA) or the Generalized Gradient Approximation (GGA) without any attempt to correct for on-site correlations with a Hubbard U have been used.~\cite{wawzynska07,johannes07,kang07,wawzynska08}

The main motivation for our work is to accurately assess the effect of electronic correlations on the electronic and magnetic properties of 2H-AgNiO$_2$, and also on intermetallic phases AgNi$_{1-x}$Co$_{x}$O$_2$ as well as their stability. In our work, we use quantum Monte Carlo (QMC) methods, specifically real-space variational Monte Carlo (VMC) and diffusion Monte Carlo (DMC). QMC methods are computationally expensive but highly-accurate stochastic wavefunction methods that fully incorporate electronic many-body effects.~\cite{Reynolds1982,foulkes01} Weak through strong electronic correlations are well described. The total energy obeys a variational principle allowing the effect of different choices for the input trail wavefunctions to be assessed. QMC methods have provided accurate ground state properties for strongly-correlated transition metal oxides, including VO$_2$ and AFM NiO.~\cite{zheng15,mitra15,trail17,shin17,kylanpaa19} 
In this study, we use QMC to obtain accurate ground state properties of 2H-AgNiO$_2$.
In addition, we study various structures of phases of the mixtures AgNi$_{1-x}$Co$_{x}$O$_2$ to assess their phase stability and electronic properties. 
Our results show that large concentrations of substitutional Co in AgNi$_{1-x}$Co$_{x}$O$_2$, $x \geq 0.33$, lead to an opening of an electronic band gap and stable formation energies. This suggests an interesting way to generate a metal-insulating transition concomitant with a magnetic transition, different from, {\it e.g.,} metal-insulating transitions in more classical correlated oxides, such as VO$_2$.~\cite{morin59,tselev10,huber16}  

\section{\label{sec:level2}Methods}
We used DMC within the fixed-node approximation as implemented in the QMCPACK code.~\cite{QMCPACK} 
Single Slater-determinant wavefunctions were used as trial wavefunctions in the QMC algorithm, with up to three-body Jastrow correlation coefficients in order to incorporate electron-ion, electron-electron, and electron-electron-ion correlations.
Cut-offs for the one- and two-body Jastrows were set as the Wigner-Seitz radius of the given supercell while a maximum of 5.0~Bohr was used as the cut-off for the three-body term. 
Single-particle orbitals in the QMC trial wavefunctions were generated by solving the Kohn-Sham equations using DFT.
All DFT calculations in this study were performed with a plane-wave basis set with a 700~Ry kinetic-energy cut-off and $8\times8\times8$ $k$-point grids using the QUANTUM ESPRESSO code.~\cite{giannozzi09}
Kohn-Sham orbitals in the Slater determinant were generated using Perdew-Burke-Ernzerhof (PBE) parametrization\cite{perdew97} of the generalized gradient approximation (GGA) exchange-correlation (XC) functional.
In order to account for on-site Coulomb interactions of strongly localized $d$ orbital in Ni, we used a Hubbard ``U'' for the Hubbard correction within the DFT+U formalism.~\cite{anisimov91,dudarev98} 
Norm-conserving pseudopotentials for Ni and O in this study were the same as used in a previous QMC study of AFM NiO.~\cite{shin17} 
The Ag and Co pseudopotentials were correlation-consistent effective-core potentials (ccECPs) wherein fully-correlated all-electron calculations -- primarily coupled-cluster calculations -- were used as references for the parameterization of the ECPs.~\cite{bennett17,bennett18,annaberdiyev18,wang19} 
Because ccECP pseudopotentials are hard-core and therefore require large kinetic-energy cut-off,  700 Ry kinetic-energy,  for Ag, we applied the hybrid orbital representation that combines a local atomic basis set and B-splines in order to reduce memory requirements of the QMC.~\cite{luo18} 
DMC calculations were done using 0.005~Ha$^{-1}$ time steps within the non-local $T$-move approximation \cite{casula10} 
In order to reduce one-body finite-size effects from the periodic boundary conditions applied in the DMC calculations, we employed twist-averaged boundary conditions\cite{lin01} with up to a maximum of 64 twists for the AgNiO$_2$ supercells.
Two-body finite-size effects were reduced using the modified periodic Coulomb interaction\cite{drummond08} and Chiesa's kinetic energy correction.~\cite{chiesa06} 
In addition to those finite-size corrections, we estimated twist-averaged DMC energies at different sizes of supercells, 48, 96, and 144 atoms cells, and extrapolated the energies to the bulk limit in order to further reduce two-body finite size effects. 

\section{\label{sec:results}Results}
\subsection{Properties of Pure 2H-AgNiO$_2$}
\begin{figure}[t]
 \includegraphics[width=4 in]{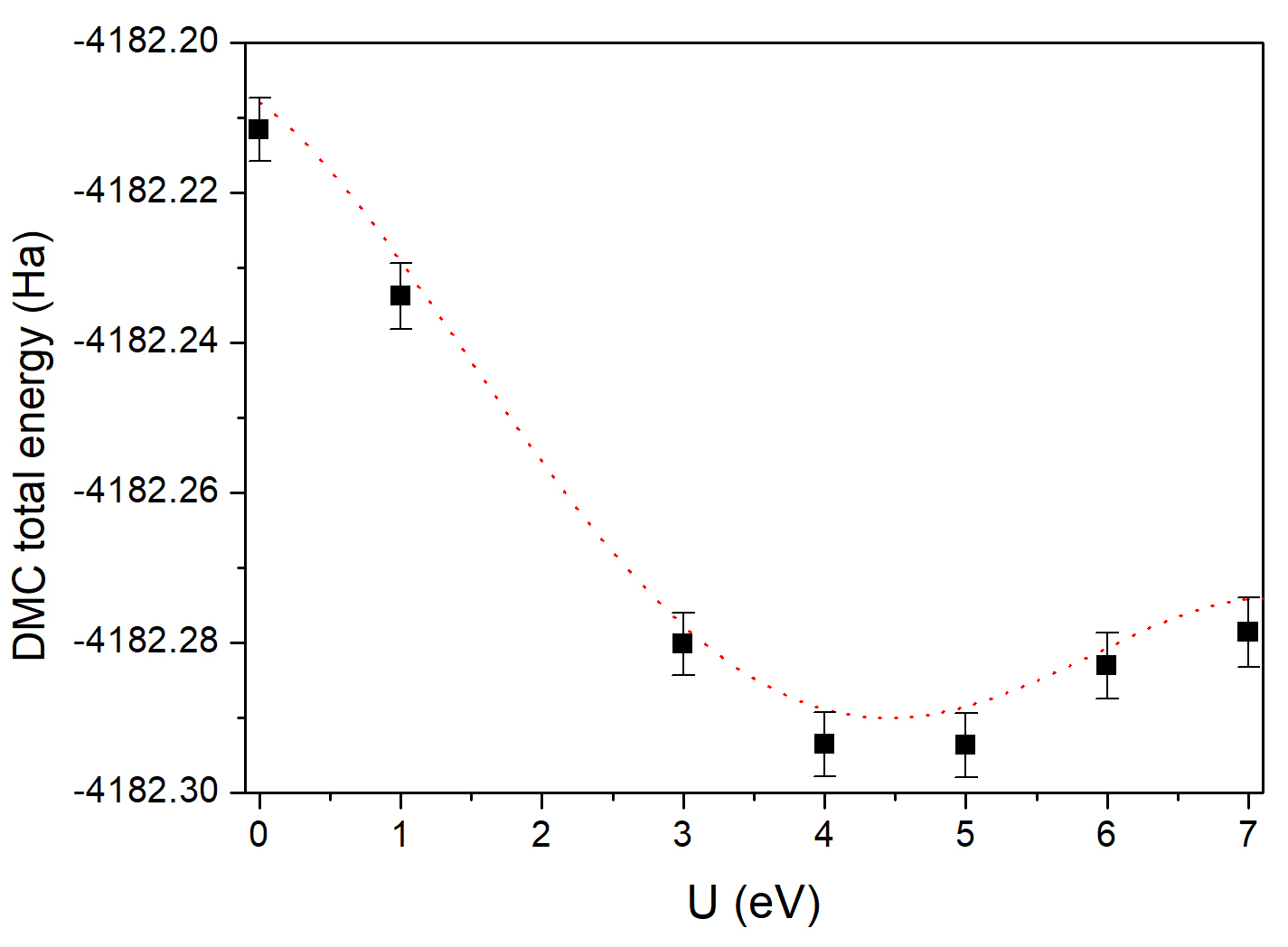}
 \caption{DMC total energy of AgNiO$_2$ as function of Hubbard U in the PBE+U trial wavefunction.}
 \label{fig:DMC_U}
\end{figure}

Previous DFT studies of AgNiO$_2$ delafossites assumed that on-site Coulomb interactions were not important and so did not use DFT+U.~\cite{wawzynska07,wawzynska08} One of the aims of our work is to examine the role of on-site Coulomb interactions in detail in order to ascertain their importance. Because there are no previously reported values for an optimal value of U, $U_{\rm opt}$, we first estimated $U_{\rm opt}$. We used a procedure established in previous works\cite{foyevtsova14,luo16,shin17,shin18,saritas19} that has proven to be a reliable and unbiased way to estimate $U_{\rm opt}$ for transition-metal oxides. In this procedure, we minimize the DMC total energy of the PBE+U trial wavefunction as a function of U. Because the DMC total energy obeys a variational principle, this energy will exhibit a minimum.  
Specifically, in DMC, the minimization of the energy with respect to U is a one-parameter optimization of the many-body wavefunction nodal surface.   
For simplicity we assume the same U value for all Ni atoms, regardless of their local charge states. 
Figure~\ref{fig:DMC_U} shows the DMC total energy for the AFM AgNiO$_2$ unit cell as function of the value of U in the PBE+U trial wavefunction.
Using a quartic fit, we estimated an optimal U value of $U_{\rm opt}=4.4(1)$~eV for Ni, which is close to the DMC $U_{opt}=4.7(2)$~eV found in AFM NiO.~\cite{shin17}  We use the value of U = 4.4 eV obtained from DMC for all subsequent PBE+U calculations in this study.

\begin{figure}[t]
 \includegraphics[width=6 in]{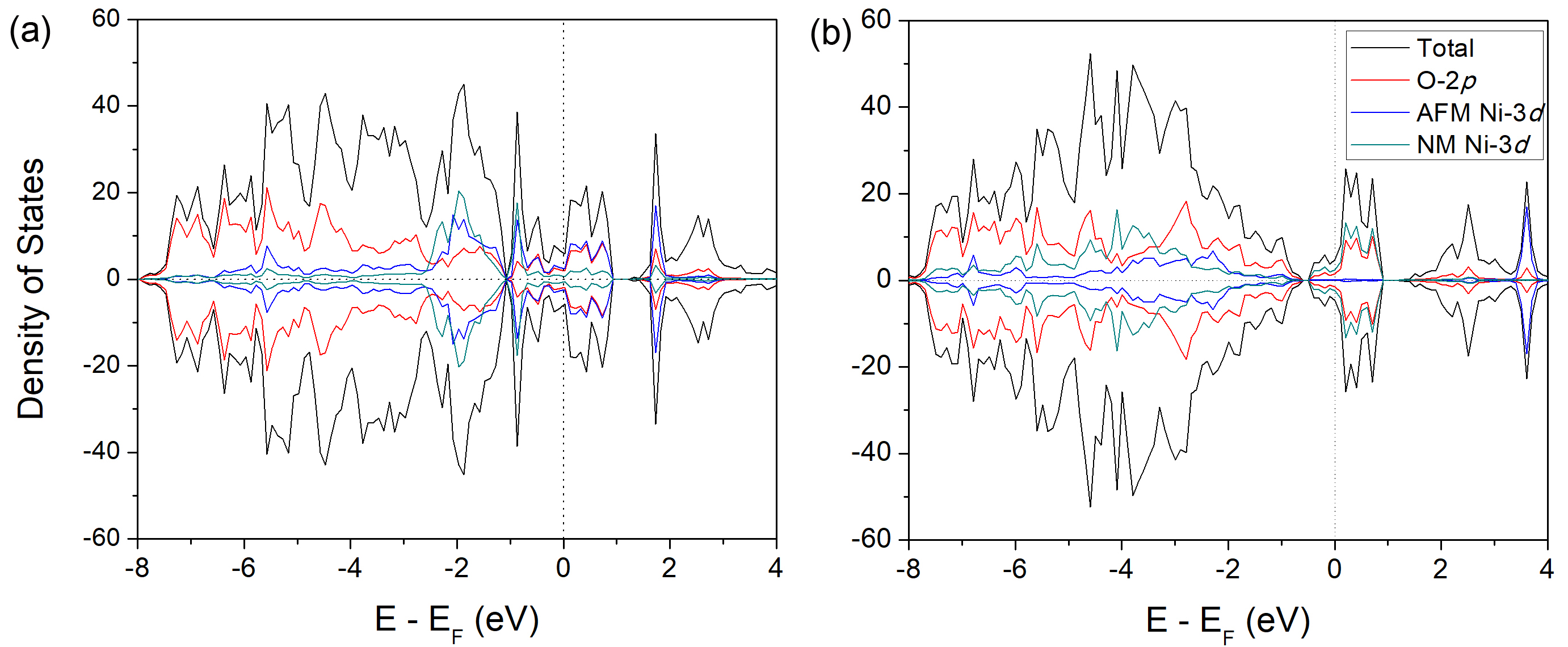}
 \caption{Projected density of states of AFM 2H-AgNiO$_2$ from (a) PBE and (b) PBE+U, with U=4.4 eV.}
 \label{fig:AgNiO2_PDOS}
\end{figure}

To investigate how varied $p$-$d$ hybridization within the DFT+U scheme may change the electronic properties of AgNiO$_2$, we first compare the electron density-of-states (DOS) obtained using PBE and PBE+U. 
As expected, the DOS for 2H-AgNiO$_2$ clearly exhibits metallic features with filled states  at the Fermi level both for PBE and PBE+U (Fig.~\ref{fig:AgNiO2_PDOS}).  
For both levels of theory, we can see a small gap beginning about 1~eV above the Fermi level, which suggests the possibility of tuning metallic 2H-AgNiO$_2$ to a semiconductor by tuning this gap to open at the Fermi level through, {\it e.g.,} hole doping. 
This gap is wider and closer to the Fermi level in PBE+U, as shown in Fig.~\ref{fig:AgNiO2_PDOS}(b).
In addition, we confirmed that the Hubbard U leads to more semimetallic electronic properties of AgNiO$_2$ as the conduction band minimum in PBE+U is closer to the Fermi level with lower DOS than in PBE. 
This suggests that localized Ni 3$d$ orbitals induce a semimetallic nature in AgNiO$_2$, and that AgNiO$_2$ possesses an intriguing potential of tuning the band gap to a semiconductor or insulator.

\begin{figure}[t]
 \includegraphics[width=6 in]{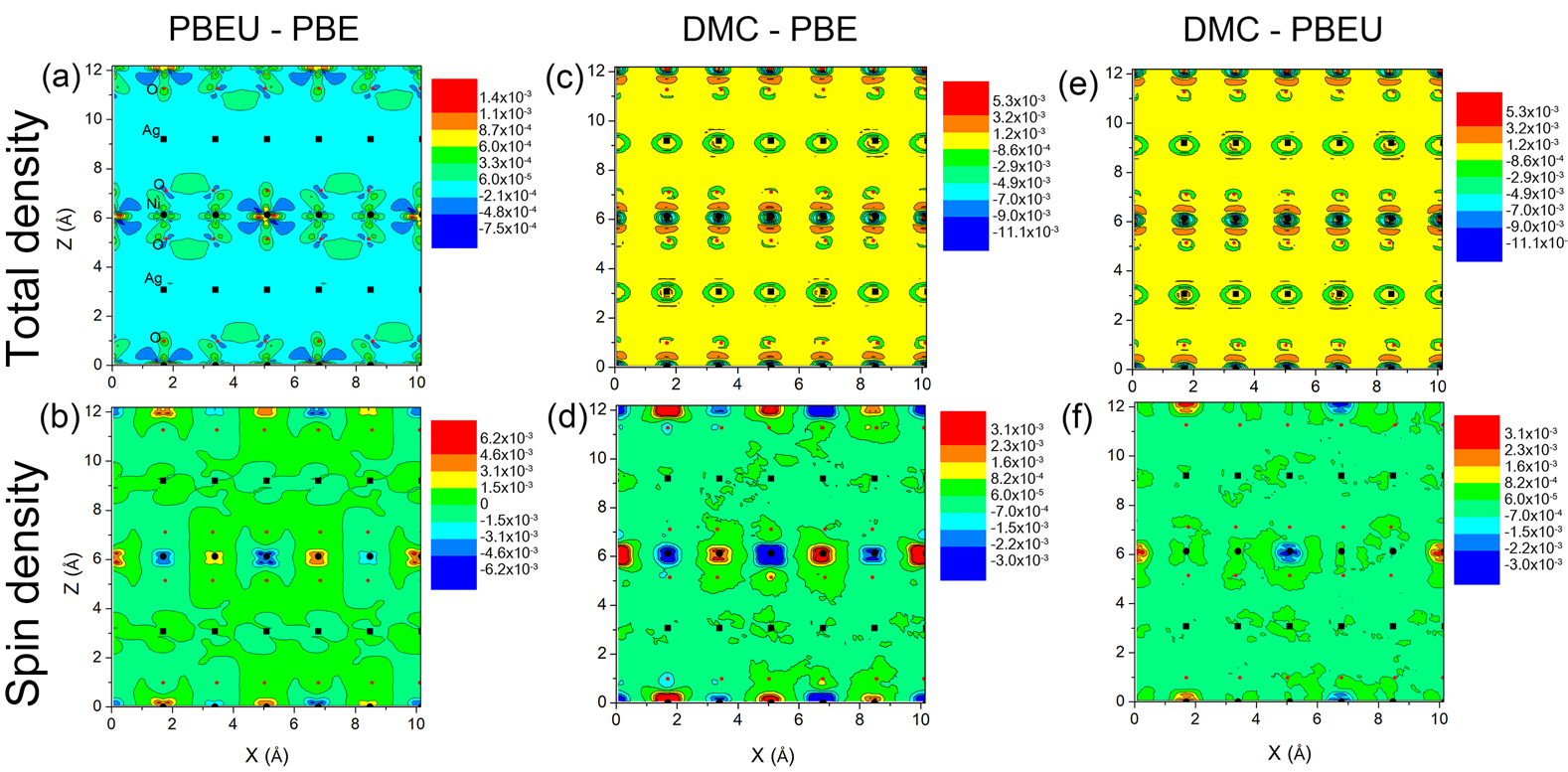}
 \caption{(a) Charge and (b) spin density difference between PBE+U and PBE of a NiO$_2$ layer of AgNiO$_2$, those between DMC and PBE ((c) and (d)), and between DMC and PBE+U ((e) and (f)). The density differences are projected onto the (010) plane and in units of \AA$^{-3}$.}
 \label{fig:AgNiO2_density}
\end{figure}

For further analyses of the effects of U on semimetallic AgNiO$_2$, we compared total charge and spin densities obtained from PBE and PBE+U. 
Figure~\ref{fig:AgNiO2_density}(a) and (b) show significant differences in both charge and spin densities  between PBE+U and PBE near the Ni sites -- accumulation and depletion can be found near the Ni sites in both the charge and spin density differences. 
The charge density differences between PBE+U and PBE induced by the Hubbard U are mainly located on the octahedral NiO$_6$ structures, with no significant changes near the Ag sites. 
Within the NiO$_2$ layers, there is a rather pronounced charge density redistribution induced by Hubbard U on the Ni-O bond. This shows that the Hubbard U strongly affects the $p$-$d$ hybridization of the Ni-O bonds, even though AgNiO$_2$ is in a semimetallic phase.  
Among the Ni sites, there is a large charge accumulation on the Ni$^{2+}$ sites (Ni2) that also possess large magnetic moments.
This indicates that there is discrepancy between the magnetic moments obtained by PBE and PBE+U,  as the Hubbard U affects the magnetic moment on Ni. 
In addition to the charge density difference, we can also see that PBE underestimates the spin  density on Ni sites relative to PBE+U (see Fig.~\ref{fig:AgNiO2_density}), which is analogous to  results obtained in an earlier DMC study of AFM NiO,\cite{shin17}, although the spin density difference is smaller for AgNiO$_2$ than for insulating NiO. 
As assumed, it is clear that influence of the Hubbard U is not as large in semimetallic AgNiO$_2$ compared to its effect in insulating NiO; however, we conclude that the existence of localized 3$d$ orbitals is still leads to moderate effects in AgNiO$_2$ because of the large density differences between PBE and PBE+U.    

In order to further accurately assess the electronic properties of 2H-AgNiO$_2$, we performed DMC calculations of AgNiO$_2$ using a PBE+U trial wavefunction with the optimal value of U. 
We estimated the cohesive energy of AgNiO$_2$ by computing $E(AgNiO_{2})-E(Ag)-E(Ni)-2E(O)$, where $E(AgNiO_{2})$, $E(Ag)$, $E(Ni)$, $E(O)$ are the DMC total energy of AgNiO$_2$ and that of atomic Ag, Ni, and O, respectively. 
The computed DMC AgNiO$_2$ cohesive energy with full incorporation of the finite-size analysis is 14.23(3)~eV/f.u., which is significantly smaller than the PBE result of 15.21~eV/f.u. but consistent with PBE+U one of 14.21~eV/f.u..
{Significantly larger PBE cohesive energy than DMC seems to be related with overestimation of NiO cohesive energy compared to corresponding experimental result within PBE functionals.\cite{shin17}
Although experimental values of the AgNiO$_2$ cohesive energy are not available to the best of our knowledge, the large differences in cohesive energy clearly shows a large discrepancy between the DMC, DFT, and DFT+U schemes in dealing with the electronic structure of AgNiO$_2$. 
The charge density difference between DMC and PBE+U, $\rho({\rm DMC})-\rho({\rm PBE+U})$, shows a charge density accumulation on Ni-O complexes in DMC relative to PBE+U, somewhat similar to the charge density difference $\rho({\rm PBE+U}) - \rho({\rm PBE})$, but the charge accumulation in $\rho({\rm DMC})-\rho({\rm PBE+U})$ is concentrated on specific Ni-O pairs in the $yz$ plane, while density difference $\rho({\rm PBE+U}) - \rho({\rm PBE})$ is more spread out over the entire NiO$_6$ layer. 
From this anisotropic density accumulation in DMC relative to PBE+U, we suspect there is a similar symmetry-breaking in the Ni-O bonds to that already seen in DMC studies of NiO and HfO$_2$.~\cite{shin17,chimata19}
In Fig.~\ref{fig:AgNiO2_density}(d) and (f), we see strong spin accumulation and depletion only on the AFM Ni sites. 
This tells us that magnetic moment on the Ni sites is significantly underestimated in both PBE and PBE+U compared to DMC. 
In order to compare the DMC and DFT magnetic moments, we computed the magnetic moments on Ni sites as function of U.
Figure~\ref{fig:DMC_mag} shows that the DFT magnetic moment increases monotonically on the AFM Ni sites Ni2 as U increases. 
However, even at large values of U, up to 6~eV where PBE+U magnetic moment shows the largest value, the PBE+U moment is still smaller than DMC magnetic moment. 
The estimated DMC magnetic moment on the AFM Ni sites is 1.71(1)~$\mu_{B}$, which is slightly larger but in the good agreement with the reported local magnetization of Ni, 1.552(7) $\mu_{B}$.~\cite{wheeler09}
We see that PBE+U magnetic moment shows empirically closest result with the experimental one in U $\sim$ 2 eV with 1.58 $\mu_{B}$ while PBE without U exhibits smaller value of 1.46 $\mu_{B}$.
From this analysis, we conclude that the Hubbard U significantly affects the band gap and magnetic moment of 2H-AgNiO$_2$, and the addition of a Hubbard U is necessary in order to  achieve reasonably accurate magnetic moment and charge density within DFT.  

\begin{figure}[t]
 \includegraphics[width=4 in]{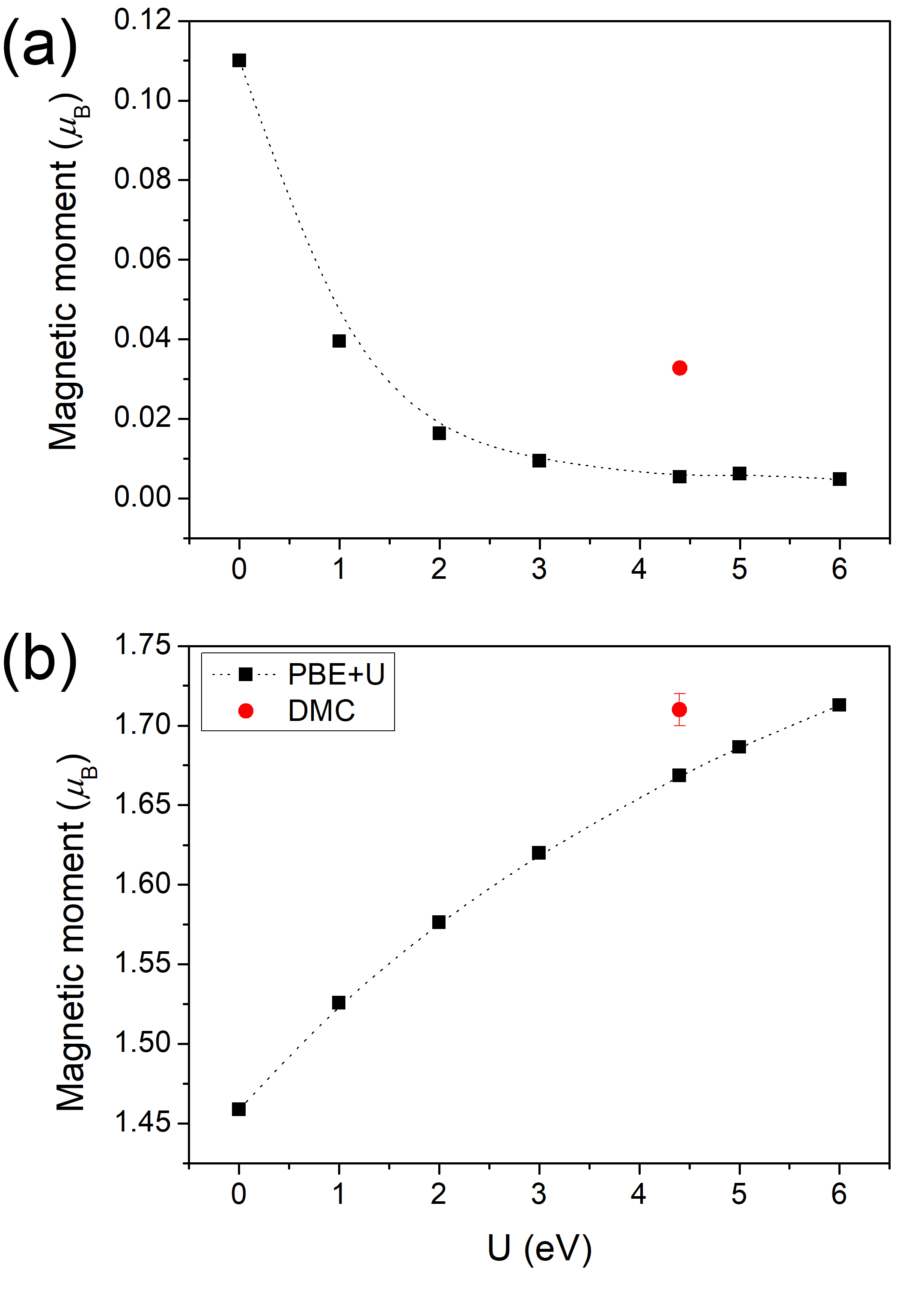}
 \caption{Magnetic moments of (a) Ni1 and (b) Ni2 sites as function of U obtained using PBE+U (black squares) and DMC (red circle).}
 \label{fig:DMC_mag}
\end{figure}

\subsection{Moderate Co doping: Metal insulator-transition for x=0.33}
\begin{figure}[t]
 \includegraphics[width=5 in]{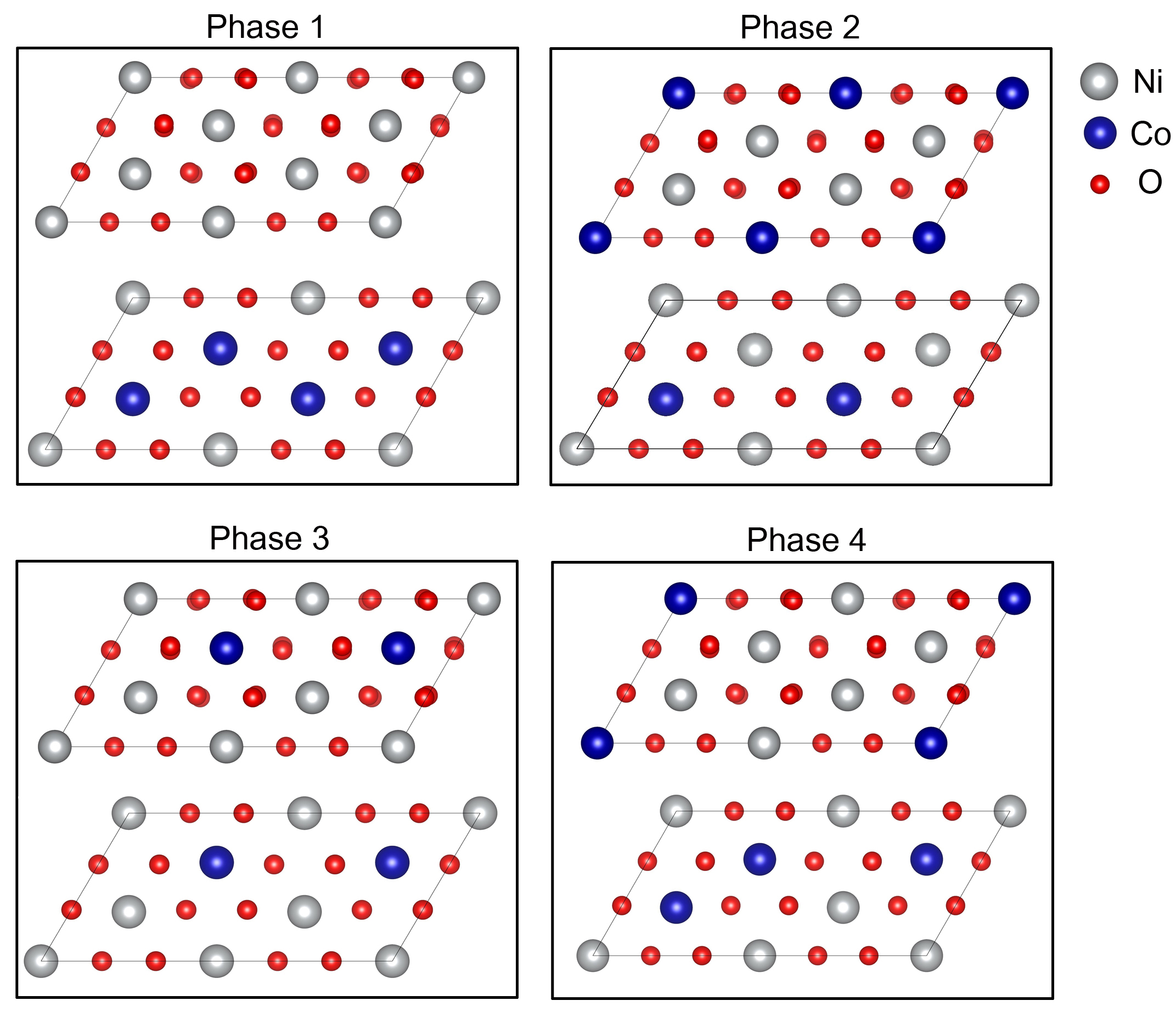}
 \caption{Upper and lower Co-doped NiO$_2$ layers for four different phases of AgNi$_{0.66}$Co$_{0.33}$O$_2$. The blue spheres indicate Co dopants.}
 \label{fig:phase}
\end{figure}
The PBE+U and DMC results for AgNiO$_2$ described in the previous section indicate the possibility of a metal-insulator transition based on the observations both in PBE+U and DMC of a small of electron density from states in the conduction band just below the Fermi level. 
This suggests that the introduction of dopants or other defects into stoichiometric AgNiO$_2$ may provide a path to move the conduction band minimum above the Fermi level. 
Various transition-metal doped delafossites have in fact been studied previously as transition-metal doping has been known to enhance $p$-type semiconductor properties. 
Among various transition metal candidates, we consider here Co as a dopant and investigate how Co doping influences the electronic properties and band gap opening in 2H-AgNiO$_2$. 
When studying these intermetallics, it is crucial first to obtain an accurate structure as the equilibrium structure and geometry vary with the concentration of dopants and with their locations, and the electronic structure in turn depends strongly on the geometry of the structure. 
We attempted to obtain a good quality lattice structure for AgNi$_{1-x}$Co$_{x}$O$_2$ by considering both the pure AgNiO$_2$ geometry but with dopants on Ni1 sites, and a fully relaxed structure within DFT+U framework.
To compare these two geometries and to choose an energetically stable geometry for the intermetallic, we estimated the DMC total energy for these structures. 
The result is that the pure 2H-AgNiO$_2$ structure with Co on the Ni1 sites exhibits a lower FN-DMC energy than the fully relaxed structure.  
Therefore, we used the pure 2H-AgNiO$_2$ as a structure for the intermetallic AgNi$_{1-x}$Co$_{x}$O$_2$.
Details in FN-energy comparison between different geometries are in Supplemental information.$^{\dag}$
In order to optimize the trial wavefunction for AgNi$_{1-x}$Co$_{x}$O$_2$, we determined an optimal U-value of 4.0(1)~eV for the Co dopants by minimizing the DMC total energy for the 2H-AgCoO$_2$ structure (see Supplemental Information).$^{\dag}$ 
Among potentially available Co concentrations of AgNi$_{1-x}$Co$_{x}$O$_2$, we first considered AgNi$_{0.66}$Co$_{0.33}$O$_2$ structure.
Although the existence of MIT can be expected on AgNi$_{0.66}$Co$_{0.33}$O$_2$ since it is experimentally reported on $x \sim 0.3$,\cite{shin93} the energetic stability of phases that result from {\it random} Co substitutions at various Co concentration is unclear.
In order to investigate the relative stability of various random phases and the dependencies of their electronic and optical properties on substitutional sites, we considered additional phases of AgNi$_{0.66}$Co$_{0.33}$O$_2$ wherein four Ni$^{+2}$ Ni1 sites out of a total of eight are replaced by Co dopants.
We did not consider substitution of Co on the AFM Ni2 sites because calculations showed that this leads to a collapse of the  magnetic order and an energetically unstable structure of the mixture.
Because there are too many AgNi$_{0.66}$Co$_{0.33}$O$_2$ configurations with four Co atoms on eight possible sites to make comprehensive DMC calculations practical,
we selected only four different phases, shown in Fig.~\ref{fig:phase}.
\begin{figure}[t]
 \includegraphics[width=5 in]{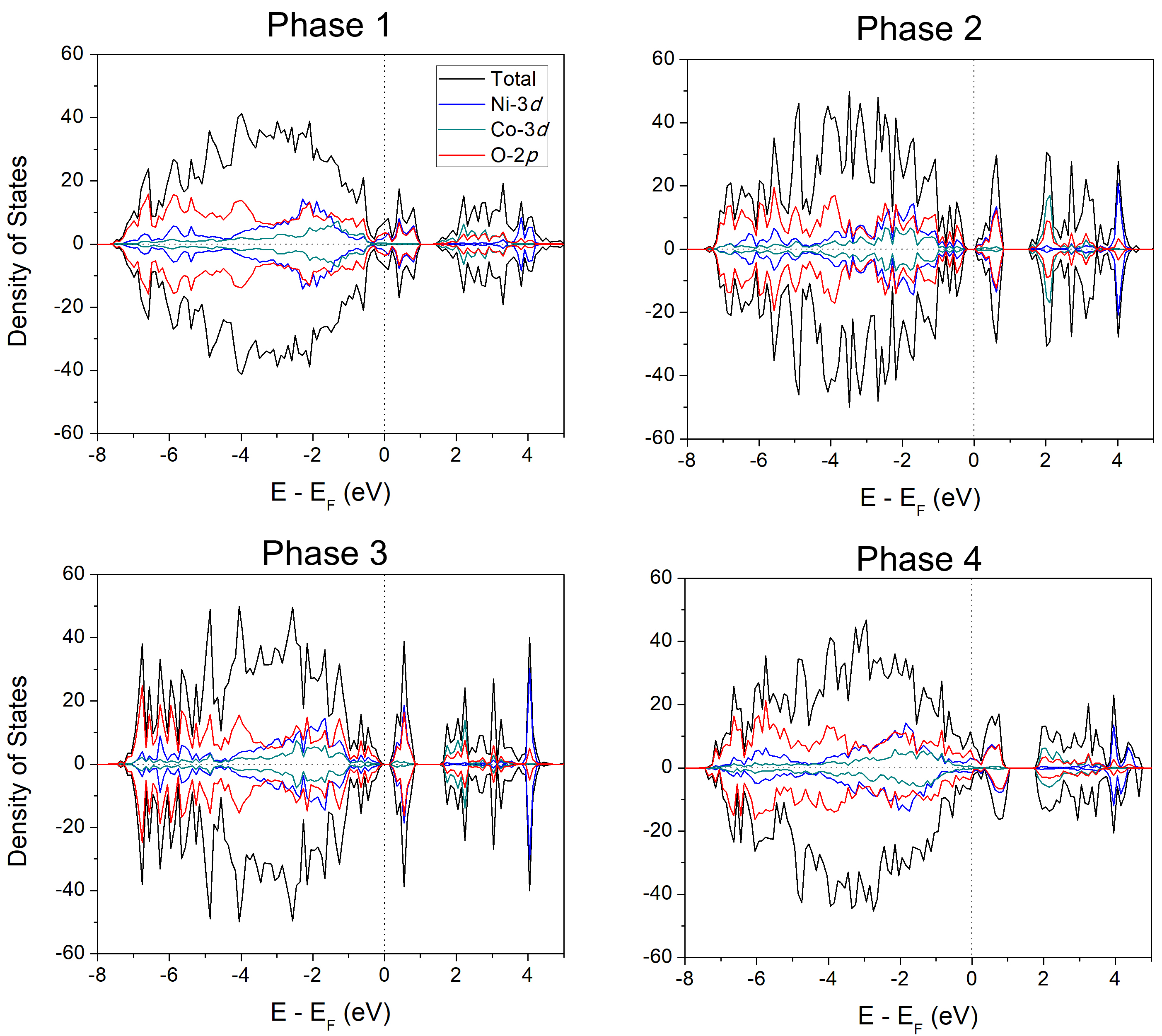}
 \caption{Projected density-of-states for four different phases of AgNi$_{0.66}$Co$_{0.33}$O$_2$.}
 \label{fig:phase_PDOS}
\end{figure}

We first compute the PBE+U density-of-states of these four phases in order to compare their optical properties. 
As can be seen in Fig.~\ref{fig:phase_PDOS}, the optical properties of the AgNi$_{0.66}$Co$_{0.33}$O$_2$ mixture depends strongly on which of the Ni1 sites are substituted with Co. 
Phases 1 and 4 show completely closed band gaps and metallic densities-of-states; however, phases 2 and 3 exhibit open band gaps. 
Because of the completely different electronic properties of the four phases, with phases 1 and 4 metallic and phase 2 and 3 semiconductor-like,  
and the very large differences in densities-of-states near the Fermi level, we conclude that the electronic and optical properties vary strongly with the specific sites used for Co-substitution, and the detailed properties of AgNi$_{0.66}$Co$_{0.33}$O$_2$ can potentially be controlled by selectively choosing the sites for substitution. 
\begin{figure}[t]
 \includegraphics[width=4 in]{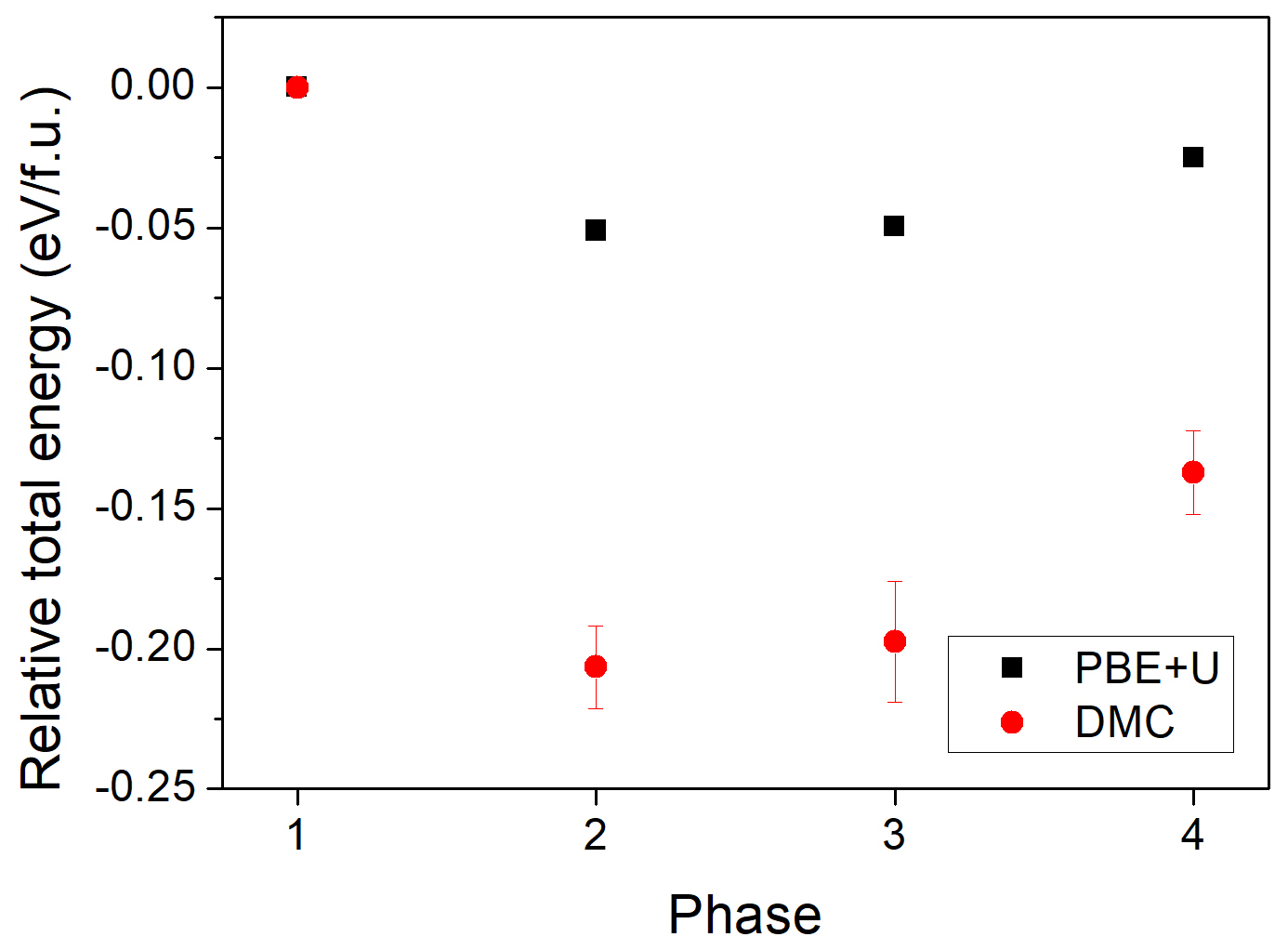}
 \caption{PBE+U and DMC relative energy between four different phases of AgNi$_{0.66}$Co$_{0.33}$O$_2$.}
 \label{fig:phase_energy}
\end{figure}

Because there are many possible metallic and semiconducting phases of
AgNi$_{0.66}$Co$_{0.33}$O$_2$, it is important to find the most stable one. We estimated the DMC total energy of four candidates based on symmetry. 
Figure~\ref{fig:phase_energy} shows the PBE+U and the DMC total energy differences between the four phases with the energy (PBE+U and DMC, respectively) of phase 1 as reference at zero total energy. 
As can be seen in the figure, the semiconducting phases 2 and 3 have lower DMC total energy than the metallic phases 1 and 4, indicating that the semiconducting phases are more energetically favored and stable than metallic ones for the AgNi$_{0.66}$Co$_{0.33}$O$_2$ mixture. 
There is a large DMC energy difference between the metallic phases 1 and 4 in DMC and a smaller PBE+U energy difference, but a relatively small energy difference between phases 2 and 3 both for PBE+U and DMC. 
The much smaller PBE+U energy difference between the metallic and semiconducting phases than the DMC energy difference, about 0.05~eV/f.u. and 0.21(1)~eV/f.u., respectively, strongly suggests that the semiconducting phases driven by Co-substitution are due to electron correlations between the Co and Ni sites, effects that are well accounted for in DMC but not as accurately in DFT or DFT+U. 
From lower PBE+U and DMC total energies on semiconducting phases than semimetallic ones confirmed the existence of MIT, transiting favored phase from semimetallic on pristine AgNiO$_2$ to semiconducting phase on AgNi$_{0.66}$Co$_{0.33}$O$_2$.
In addition, since coexistence of semimetallic and semiconducting phase is observed at the concentration of $x = 0.33$, we assume that the critical Co concentration of MIT is located nearby $x = 0.33$, which is consistent with the experimental measurement of MIT on $x = 0.3$.~\cite{shin93}
On the other hand, we see the metallic phase in higher Co concentration on the single NiO$_6$ layer than $x = 0.33$ as seen in both phase 1 and 4. 
Since the varied structure are in-layer density fluctuations and each of these contain a layer at higher Co concentration, these results leads us to suspect the existence of reentrant phase to the metallic phase on high Co concentration over $x = 0.33$.

\subsection{High Co doping: AgNi$_{0.33}$Co$_{0.66}$O$_2$}
In order to investigate optical properties on high Co concentration for AgNi$_{1-x}$Co$_{x}$O$_2$, we additionally considered high Co concentration of $x = 0.66$, substituting all non-magnetic Ni1 sites in a hexagonal pattern (see Fig.~\ref{fig:mixture_06}).
This $x = 0.66$ seems a hypothetical structure at a concentration where no experimental result for the electronic and optical properties has been reported.
\begin{figure}[t]
 \includegraphics[width=5 in]{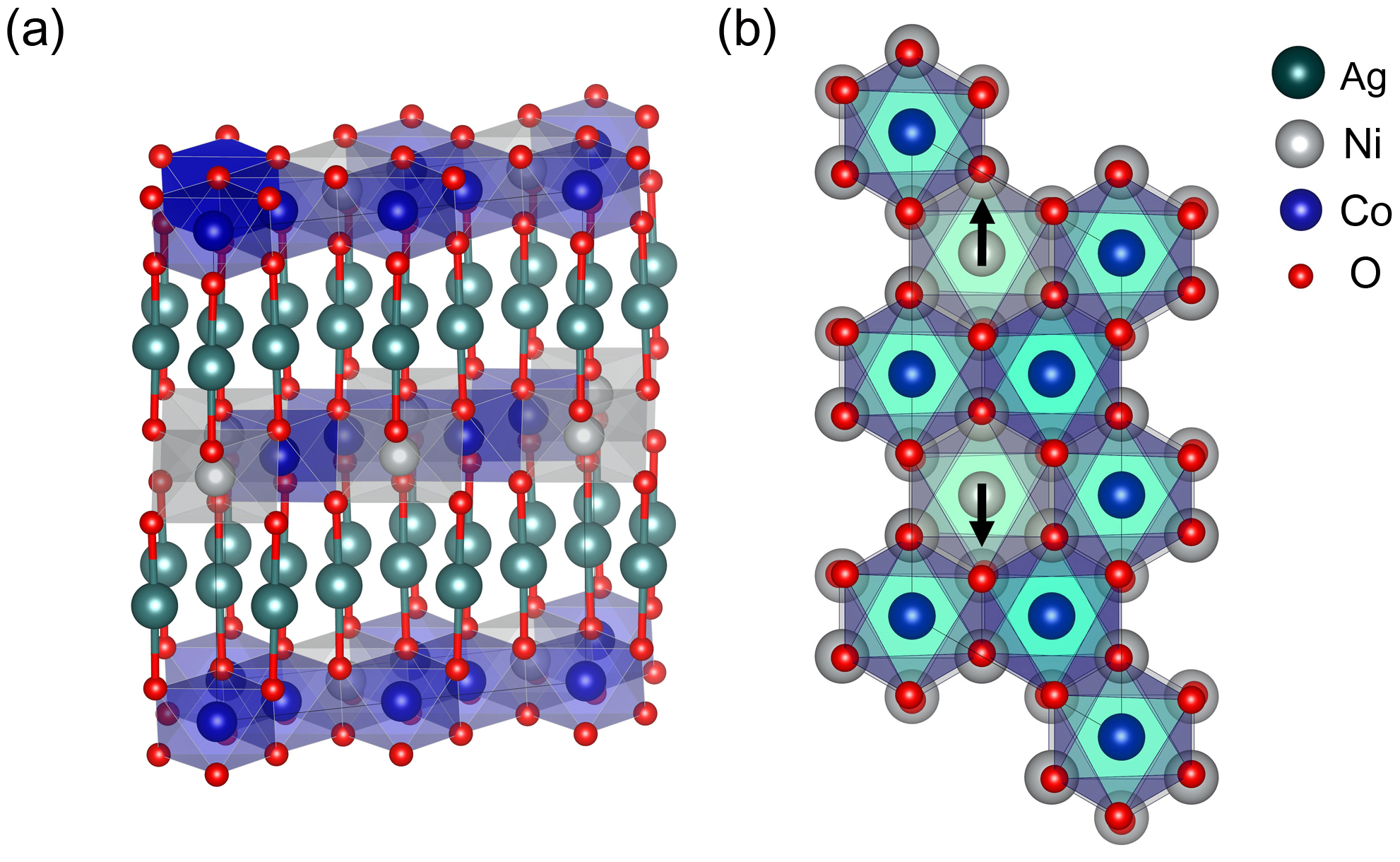}
 \caption{(a) Side and (b) top view of AgNi$_{0.33}$Co$_{0.66}$O$_2$ cell. Blue atoms represent Co.}
 \label{fig:mixture_06}
\end{figure}

The PBE+U density-of-states of AgNi$_{0.33}$Co$_{0.66}$O$_2$ (see Fig.~\ref{fig:mixture_PDOS}(a)) shows that Co-doping moves the valence band edge very close to the Fermi level, and at the conduction band edge, 3$d$-Co states have fully replaced $d$-Ni ones. 
Although the valence band edge still lies above Fermi level, the closeness of the band edge to the Fermi levels suggests that Co-substitution on the Ni$^{+3.5}$ Ni1 sites results in the electronic properties of AgNi$_{0.33}$Co$_{0.66}$O$_2$ moving from those of a semimetal closer to those of an insulator. 
\begin{figure}[t]
 \includegraphics[width=6 in]{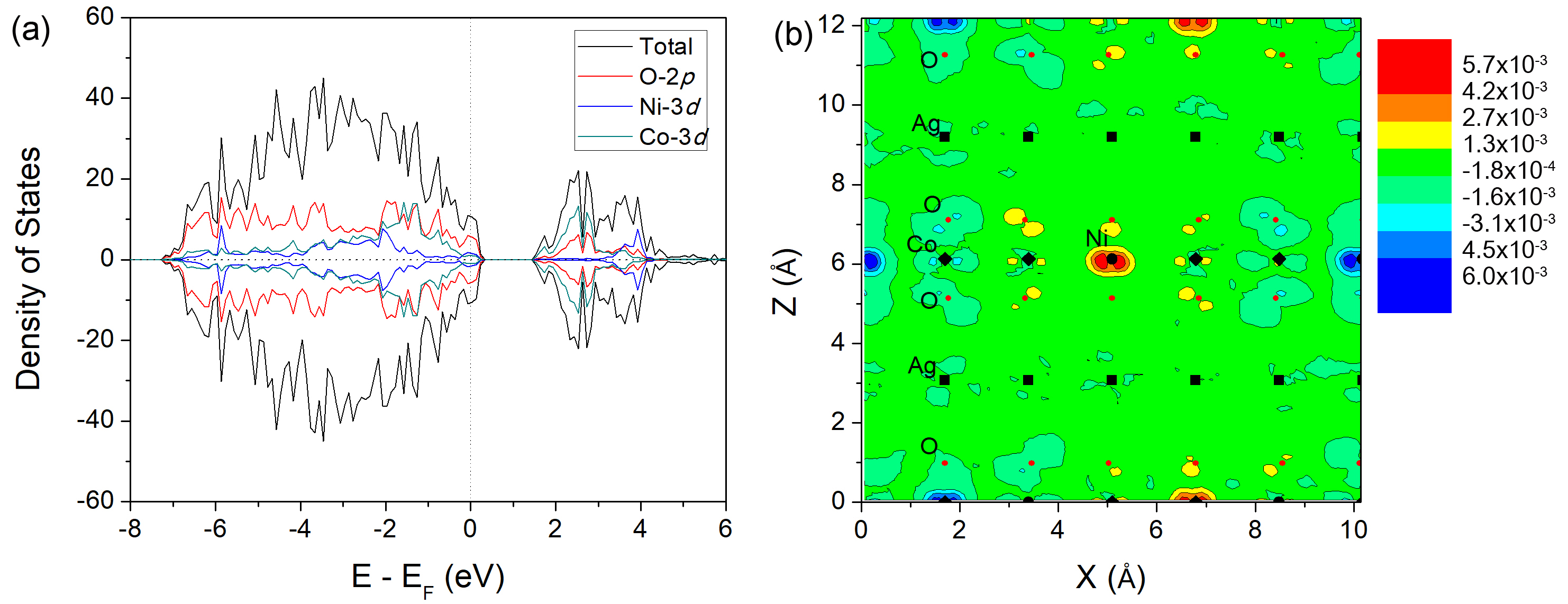}
 \caption{(a) PBE+U projected density-of-states of AgNi$_{0.33}$Co$_{0.66}$O$_2$, and (b) DMC spin density difference between AgNiO$_2$ and AgNi$_{0.33}$Co$_{0.66}$O$_2$.}
 \label{fig:mixture_PDOS}
\end{figure}

Figure~\ref{fig:AgNiO2_PDOS}(b) shows the DMC spin density difference between AgNi$_{0.33}$Co$_{0.66}$O$_2$ and AgNiO$_2$. The figure shows a density changes on the AFM Ni sites Ni2, but density change is opposite in sign to the induced AFM magnetic moments. 
This tells us that the magnetic moments on Ni sites on AgNi$_{0.33}$Co$_{0.66}$O$_2$ are smaller than in pristine AgNiO$_2$, and this is confirmed by a DMC estimate of the magnetic moment if 1.60(2)~$\mu_{b}$ for AgNi$_{0.33}$Co$_{0.66}$O$_2$, which is smaller than the moment of 1.81~$\mu_{b}$ for the same Ni2 sites in AgNiO$_2$.
These results suggest that reentrance to the metallic phase can be possible at high Co doping.
This behavior is also consistent with phase 1 and 4 in Fig.~\ref{fig:phase} for AgNi$_{0.66}$Co$_{0.33}$O$_2$, and confirms that reentrance to the metallic phase from insulator can be possible in high Co concentration.

\subsection{Stability of AgNi$_{1-x}$Co$_{x}$O$_2$}
\begin{figure}[t]
 \includegraphics[width=5 in]{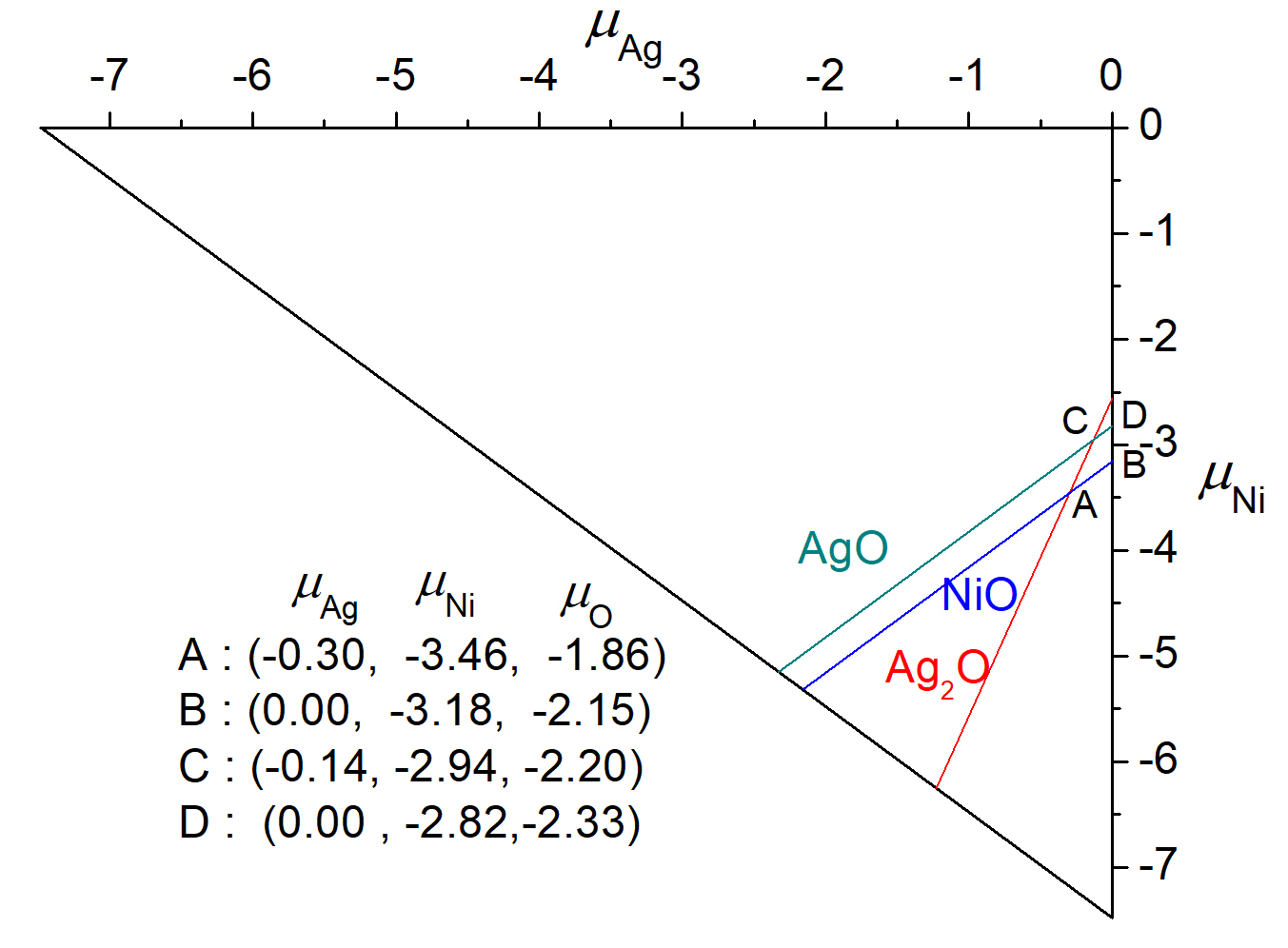}
 \caption{Illustration of the accessible chemical potential range for AgNiO$_2$ from PBE+U.}
 \label{fig:triangle}
\end{figure}
The thermodynamic stability of the AgNiO$_2$ delafossite and its doped variants depends  strongly on their formation energies. 
In order to estimate optimal conditions for formation of  AgNi$_{1-x}$Co$_{x}$O$_2$, we calculated the enthalpy of formation for AgNiO$_2$ and its doped variants under different growth conditions.
The enthalpy of formation of AgNiO$_2$ can be estimated by computing $\Delta H^{AgNiO_{2}}_{f} = \Delta\mu_{Ag} + \Delta\mu_{Ni} + 2\Delta\mu_{O}$ where $\mu_X$ indicates the chemical potential for given atom $X$.
In order to prevent formation of competing phases and phase separation or decomposition of AgNiO$_2$, these chemical potentials should be constrained as follows:
\begin{equation}
  \begin{aligned}
 \Delta\mu_{Ag} + \Delta\mu_{O} \leq \Delta H^{AgO}_{f}\\
 2\Delta\mu_{Ag} + \Delta\mu_{O} \leq \Delta H^{Ag_{2}O}_{f}\\
 \Delta\mu_{Ni} + \Delta\mu_{O} \leq \Delta H^{NiO}_{f}\\
 3\Delta\mu_{Co} + 4\Delta\mu_{O} \leq \Delta H^{Co_{3}O_{4}}_{f}.\\
   \end{aligned}  
\end{equation}
One phase boundary to AFM Co$_3$O$_4$ is avoided by respecting a constraint on Co-doping in AgNi$_{1-x}$Co$_{x}$O$_2$. 
We computed the enthalpy of formation for AgNiO$_2$, its decompositions, and Co$_3$O$_4$ using DFT+U reference energies for solid Ag, Ni, Co, and for gas-phase molecular O$_2$. 
We did not try to estimate DMC formation energies because of previously reported difficulties in computing an accurate reference energy for ferromagnetic bulk Ni using a single Slater-determinant trial wavefunction.~\cite{shin17}
\begin{table}[t]
\small
\centering
\caption{PBE+U formation energies in eV for AgNi$_{1-x}$Co$_{x}$O$_2$ under different growth conditions.}
\label{tab:PBE_formation}
\begin{tabular}{c|c|cccc}
\hline \hline
                        &  \multirow{2}{*}{AgNi$_{0.33}$Co$_{0.66}$O$_2$}  &  \multicolumn{4}{c}{AgNi$_{0.66}$Co$_{0.33}$O$_2$} \\
                        &            &   Phase 1      & Phase 2    & Phase 3    & Phase 4  \\ \hline
 Ag-poor:Ni-poor:O-rich &   -0.13    &    -0.97     &  -1.02    &  -1.02    &  -0.99       \\ 
 Ag-rich:Ni-poor:O-poor &   -0.16     &     -0.98    &  -1.03    &  -1.03    &  -1.00   \\ 
 Ag-poor:Ni-rich:O-rich &   -0.13     &  -0.97        &  -1.02     &  -1.02     &  -0.99  \\ 
 Ag-rich:Ni-rich-O-poor &   -0.15     &  -0.98        &  -1.02     &  -1.03    &  -1.00    \\ \hline \hline
\end{tabular}
\end{table}

Based on the computed $\Delta H_{f}$ using PBE+U, the phase diagram of AgNiO$_2$ can be illustrated as a function of the allowed ranges of the chemical potentials of Ag, Ni, and O,  given the constraints on them, as shown in Fig.~\ref{fig:triangle}.~\cite{persson05,walsh08} 
Within the boundaries given by the constraints on the enthalpy of formation of AgO, Ag$_2$O, and NiO, we obtained the following chemical potentials of ($\Delta\mu_{Ag},\Delta\mu_{Ni},\Delta\mu_{O}$) under different growth conditions: Ag-poor:Ni-poor:O-rich A(-0.10, -2.85, -0.05), Ag-rich:Ni-poor:O-poor B(0.00, -2.75, -0.15), Ag-poor:Ni-rich:O-rich C(-0.08,-2.78,-0.10), and Ag-rich:Ni-rich:O-poor D(0.00, -2.71, -0.17).
Using these chemical potentials, the formation energy of a Co defect is given by
\begin{equation}
E_{f}(Co)= E_{AgNi_{1-x}Co_{x}O_{2}} - E_{AgNiO_{2}} + \Sigma_{i}n_{i}(E_{i}+\Delta\mu_{i}),   
\end{equation}
where $E_{AgNi_{1-x}Co_{x}O_{2}}$ $E_{AgNiO_{2}}$, $n_{i}$, and $E_{i}$ are the total energy of AgNi$_{1-x}$Co$_{x}$O$_2$ and AgNiO$_2$, the number of added ($n_{i} < 0$) and removed ($n_{i} > 0$) atoms for substitutions, and the reference energy from the standard solid or gas-phase reference states for the constituent elements, respectively.   
\begin{table}[t]
\small
\caption{PBE+U formation energies in eV for AgNi$_{1-x}$Co$_{x}$O$_2$ against elemental solids (gaseous oxygen) and binary oxides under stoichiometric conditions.}
\label{tab:PBE_formation_binary}
\begin{tabular}{c|c|cccc}
\hline \hline
                 &         \multirow{2}{*}{x = 0.66}  &  \multicolumn{4}{c}{x = 0.33} \\
                                    &    & Phase 1 & Phase 2    & Phase 3    & Phase 4  \\ \hline
Elemental  &  -3.58    &    -4.22     &  -4.27    &  -4.27    &  -4.25       \\ 
Binary  &  -0.09    &    -0.87     &  -0.92    &  -0.92    &  -0.89       \\ \hline \hline
\end{tabular}
\end{table}

Table~\ref{tab:PBE_formation} summarizes the computed formation energies of Co dopants in AgNi$_{0.66}$Co$_{0.33}$O$_2$ and AgNi$_{0.33}$Co$_{0.66}$O$_2$.
As is seen in the Table, PBE+U predicts spontaneous formation of Co-defects for all phases and growth conditions for which the formation energy of the defect is negative. It has previous been reported that PBE+U tends to underestimate the formation energy of defects in transition metal oxide systems~\cite{santana15,shin17,chimata19,ichibha23}, so spontaneous defect formation may not occur. The PBE+U results lead us to confirm that the Ag-rich:Ni-poor:O-poor growth condition is the most favorable one for pure AgNiO$_2$ and AgNi$_{1-x}$Co$_{x}$O$_2$ with the lowest formation energy within the given constraints.
Although PBE+U does not provide quantitatively accurate formation energies for Co doping, a qualitative comparison between various growth conditions does give guidelines for the best growth conditions for synthesizing AgNiO$_2$ and AgNi$_{1-x}$Co$_{x}$O$_2$.     

In order to compare stability of AgNi$_{1-x}$Co$_{x}$O$_2$ with the binary oxides, we compute PBE+U formation energy against elemental solids and binary oxides under stoichiometric conditions. 
In Table~\ref{tab:PBE_formation_binary}, we see large formation energy gap of $\sim$ 8 eV between one relative with elemental solids and the binaries.
With comparison of formation energies table~\ref{tab:PBE_formation} and table~\ref{tab:PBE_formation_binary}, we see that formation energies of AgNi$_{1-x}$Co$_{x}$O$_2$ under the chemical potential constraints are significantly closer to formation energies from the binary oxides than those from the elemental solids in table~\ref{tab:PBE_formation_binary}, which tells us formation of AgNi$_{1-x}$Co$_{x}$O$_2$ is almost energetically consistent with the ideal formation against binary oxides.
In addition, smaller formation energies in x = 0.66 than x = 0.33 in all growth conditions lead us to conclude relative difficulty of AgNi$_{0.33}$Co$_{0.66}$O$_2$ synthesis.

\section{\label{sec:summary} Conclusions}
We have performed DMC calculations on the AFM AgNiO$_2$ delafossite in order to obtain accurate electronic properties and magnetic moments. 
We found that the addition of Hubbard U to the DFT scheme dramatically changes the electronic and magnetic properties of AgNiO$_2$.
Using DFT+U with the U value optimized using DMC, we confirmed that AgNiO$_2$ has a  semimetallic nature induced by strong $p$-$d$ hybridization in AFM NiO$_2$ layer.
Our PBE+U and DMC studies of AgNi$_{1-x}$Co$_{x}$O$_2$ shows a metal-insulator transition at $x \sim 0.33$ by Co substitution on the non-magnetic Ni1 sites, which is in good agreement with the experimental result.
In addition to the semiconducting phase in AgNi$_{0.66}$Co$_{0.33}$O$_2$, it is found that the coexistence of metallic phase when more than $x = 0.33$ of Co dopant is substituted in the single layer of NiO$_{2}$ in AgNi$_{0.66}$Co$_{0.33}$O$_2$, leading to possible existence of the reentrance of metallic phase in high Co concentration.
This reentrant behavior in AgNi$_{1-x}$Co$_{x}$O$_2$ is confirmed in high Co concentration of AgNi$_{0.33}$Co$_{0.66}$O$_2$ where semimetallic nature is discovered in the PBE+U result.
PBE+U formation energies revealed that AgNi$_{0.66}$Co$_{0.33}$O$_2$ possesses the lowest formation energy under oxygen-rich conditions, which suggests a path for synthesizing the  AgNi$_{0.66}$Co$_{0.33}$O$_2$ mixture.
Our works clearly shows the difficulty in using PBE+U to quantitatively estimate formation energies of Co-doping in AgNiO$_2$ because of the poor description of the 3$d$ orbitals in Ni and Co. More accurate electronic structure methods than existing DFT approximations are needed for accurate prediction of electronic and magnetic properties of defective delafossites. 

\begin{acknowledgements}
This work was supported by the U.S. Department of Energy, Office of Science,
Basic Energy Sciences, Materials Sciences and Engineering Division, as part of
the Computational Materials Sciences Program and Center for Predictive
Simulation of Functional Materials. An award of computer time was provided by
the Innovative and Novel Computational Impact on Theory and Experiment (INCITE)
program. This research used resources of the Argonne Leadership Computing
Facility, which is a DOE Office of Science User Facility supported under
contract DE-AC02-06CH11357 and resources of the Oak Ridge Leadership Computing
Facility, which is a DOE Office of Science User Facility supported under
Contract DE-AC05-00OR22725. 
\end{acknowledgements}

\appendix
\section{Geometry for AgNi$_{1-x}$Co$_{x}$O$_{2}$}
\begin{figure}[t]
 \centering
 \includegraphics[width=4 in]{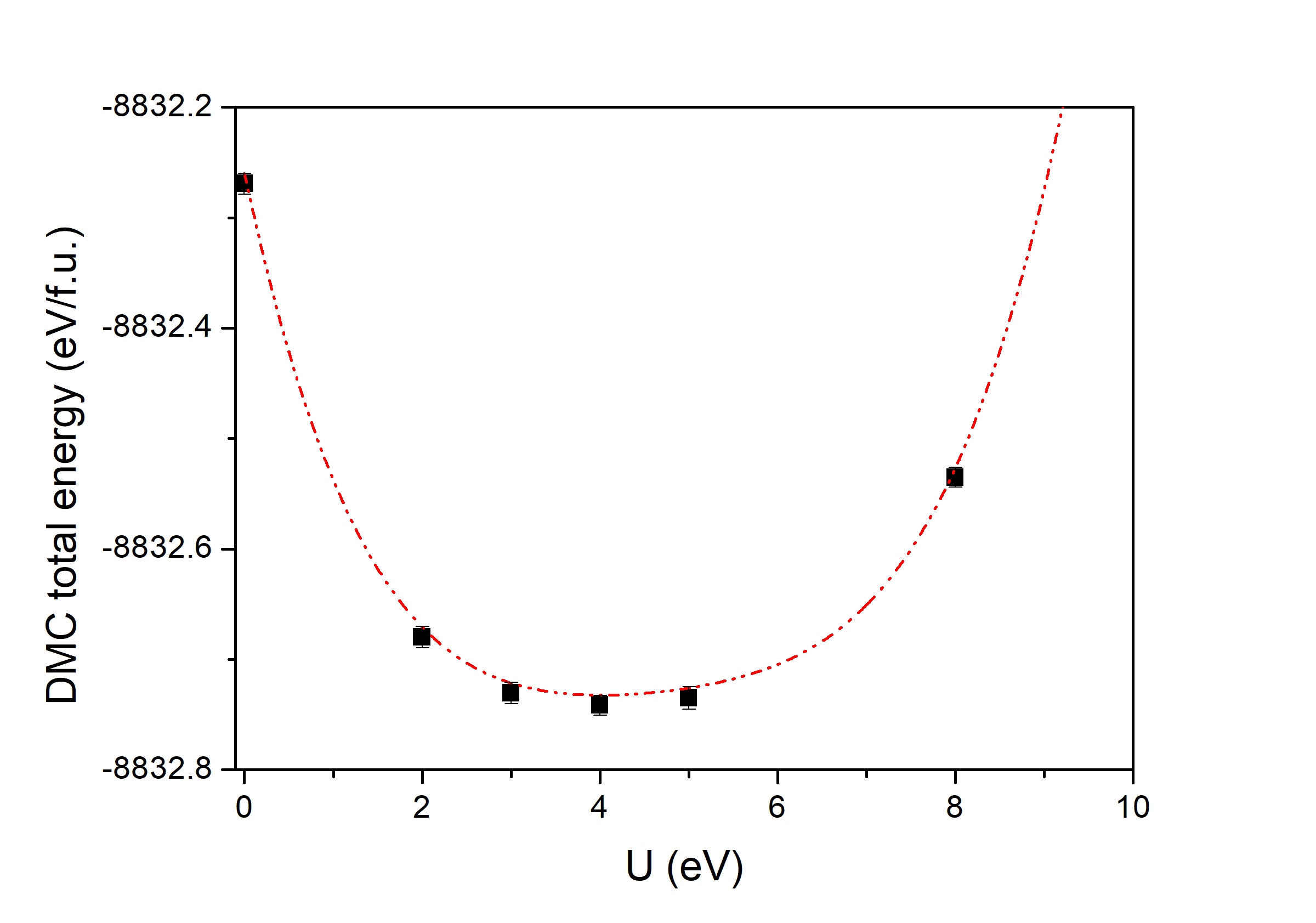}
 \caption{DMC total energy of AgCoO$_2$ as function of Hubbard U in the PBE+U trial wavefunction..} 
 \label{fig:AgCoO2_U}
\end{figure}
Obtaining accurate geometry is important in order to accurately estimate the electronic properties of doped systems.
Because detailed geometry information is not available for AgNi$_{1-x}$Co$_{x}$O$_{2}$, we compare two different geometries in order to select one to use. %to decide which one will use in this study.
A first geometry was obtained from full relaxation using PBE+U with ultra-soft
pseudopotentials, and a second one was pristine stoichiometric AgNiO$_2$
geometry with Ni1 atoms simply replaced with Co for AgNi$_{1-x}$Co$_{x}$O$_{2}$.
We performed DMC total energy calculations on these two geometries in order to
find the more  stable structure with lower fixed-node DMC energy. DMC
calculations were done with a 580 electrons supercell and the estimated DMC
energy for the DFT-relaxed AgNi$_{1-x}$Co$_{x}$O$_{2}$ geometry and for the pure
AgNiO$_2$ geometry are -9049.547(11) eV/f.u. and -9049.569(10) eV/f.u.,
respectively. It is interesting, and perhaps surprising, that the pure AgNiO$_2$
geometry has a lower fixed-node DMC energy, which implies that the AgNiO$_2$
structure is more stable than the PBE+U-relaxed AgNi$_{1-x}$Co$_{x}$O$_{2}$
structure. More disconcerting is the fact that this exercise tells us that PBE+U
does not produce a  well-optimized geometry for AgNi$_{1-x}$Co$_{x}$O$_{2}$.
Therefore, it is necessary to develop and use other geometry optimization
methods for the delafossites. That will be the subject of future work; for the
present work, we decided to use the pure AgNiO$_2$ structure for
AgNi$_{1-x}$Co$_{x}$O$_{2}$. 
\section{U scanning for AgCoO$_2$}
The optimal value of U for a Co atom in AgNi$_{1-x}$Co$_{x}$O$_2$ was chosen by scanning value of U in 2H-AgCoO$_2$ structure.
DMC calculation for scanning U was performed in 576 electrons cell within PBE+U trial wavefunction.   
As seen in Figure~\ref{fig:AgCoO2_U}, the optimal value of U for a Co atom was obtain through a quartic fit to DMC total energies and were estimated to be 4.0(1)~eV.

\bibliography{main}

\end{document}